\documentstyle[latexsym,amssymb,epsfig,aps,floats,eqsecnum,
preprint,amsfonts]{revtex}

\def\laq{\raise 0.4ex\hbox{$<$}\kern -0.8em\lower 0.62ex\hbox{$\sim$}}
\def\gaq{\raise 0.4ex\hbox{$>$}\kern -0.7em\lower 0.62ex\hbox{$\sim$}}

\font\tenbb=msbm10
\font\sevenbb=msbm7
\font\fivebb=msbm5
\newfam\bbfam
\textfont\bbfam=\tenbb \scriptfont\bbfam=\sevenbb
\scriptscriptfont\bbfam=\fivebb

\newcommand{\vphi}{\varphi} 
\newcommand{\beq}{\begin{equation}} 
\newcommand{\eeq}{\end{equation}}
\newcommand{\bea}{\begin{eqnarray}} 
\newcommand{\eea}{\end{eqnarray}}
\newcommand{\beam}{\begin{mathletters}} 
\newcommand{\eeam}{\end{mathletters}}

\newcommand{\hk}{{\widehat{k}}}
\newcommand{\ho}{{\widehat{\omega}}}

\newcommand{\hH}{{\widehat{H}}}
\newcommand{\og}{{\overline{g}}}
\newcommand{\bk}{\mbox{\boldmath${k}$}}
\newcommand{\bx}{\mbox{\boldmath${x}$}}

\begin{document}
\draft
\preprint{\vbox{\baselineskip=12pt
\rightline{IHES/P/01/05} 
\rightline{GRP/01/557}}}

\title{\large\bf The fate of classical tensor inhomogeneities \\
in pre-big-bang string cosmology}
\author{ Alessandra Buonanno${}^{a}$ and Thibault Damour${}^{b}$}
\address{$^a$ {\it Theoretical Astrophysics and Relativity Group \\
California Institute of Technology, Pasadena, CA 91125, USA }\\
{$^b$ {\it Institut des Hautes Etudes Scientifiques, 91440
Bures-sur-Yvette, France}}}
\maketitle
\begin{abstract}
In pre-big-bang string cosmology one uses a phase of 
dilaton-driven inflation to stretch an initial (microscopic)
spatial patch to the (much larger) size of the big-bang fireball.
We show that the dilaton-driven inflationary phase does not naturally
iron out  the initial classical tensor inhomogeneities unless the initial
value of the string coupling is smaller than $g_{\rm in} \alt 10^{-35}$.
\end{abstract}
\newpage
\section{Introduction}
\label{sec1}

The pre-big-bang (PBB) scenario \cite{PBB} is an attempt to use
the kinetic energy of the string-theory dilaton to drive a period of
inflation of the universe. The basic motivations of the PBB scenario
are: (i) the existence of exact (spatially homogeneous) 
dilaton-driven inflationary solutions
following from the T-duality symmetries of string-theory \cite{Ven},
and (ii) the need to bypass the fact that a tree-level dilaton 
essentially destroys \cite{CLO} the usual (potential-driven)
inflationary mechanism. In the ``stochastic'' version of the PBB 
scenario \cite{BDV99} one envisages the birth of an ensemble of 
pre-big-bang bubbles from the gravitational instability of a {\em generic}
string vacuum made of a stochastic bath of classical incoming gravitational 
and dilatonic waves. In this approach the only needed condition for the
blistering (in string units) of a PBB bubble (of size $H_{\rm in}^{-1}$,
where $H_{\rm in}$ is the initial Hubble expansion rate of a patch of space)
is similar to the corresponding condition in ``chaotic'' inflation 
\cite{Linde} (see below). Namely, locally, the inhomogeneous contributions 
(of wavelengths smaller than $
H_{\rm in}^{-1}$) to the local Friedmann equation should be fractionally
smallish (say by a factor 5) compared to the homogeneous contribution 
$ {\dot {\vphi}}^2_{\rm in}\sim H_{\rm in}^{2}$.
This ``stochastic'' PBB approach, together with 
other studies of inhomogeneous versions of PBB \cite{V97,BMUV98,crit}, was intended
to answer (or at least  to soothe) the 
concerns about fine-tuning \cite{TW97,LKB00} in the PBB
scenario. However, as far as we are aware, no complete study of the
effectiveness of the  PBB dilaton-driven inflation (DDI) in smoothing 
out initial homogeneities has been performed.
 [Note that this smoothing out of {\em classical} inhomogeneities
is the {\em prerequisite} for the discussion of the irrepressible 
{\em quantum} fluctuations that might be the seed of the large-scale
structure of the universe.] 
 Ref.~\cite{BGGMV95}
discussed the fate of (quantum) inhomogeneities during the DDI phase and
concluded that their growth, when they get out of the horizon, was
only logarithmic, but they did not analyze 
the smoothing properties of the entire pre-big-bang plus post-big-bang scenario. 
The recent 
discovery of the generic appearance of an inhomogeneous chaos, ultimately
leading to a string-scale foam near a big-crunch \cite{DH}, prompted us
to reexamine in detail the fate of initial classical inhomogeneities
during the entire evolution of the simplest PBB scenario 
(comprising an initial DDI phase matched onto a subsequent ordinary
big-bang evolution). 

In this paper we consider the ``stochastic'' version of the PBB scenario,
and study the evolution of the tensor inhomogeneities present in a {\em generic}
PBB inflationary bubble.
Our conclusions is that the PBB scenario is not very  effective  
 in smoothing out initial classical inhomogeneities (we limit ourselves to
inhomogeneities small enough for not developing into a turbulent chaos
{\em before} reaching the string scale). Indeed, analyzing tensor 
inhomogeneities,  we find that they need to be initially
unnaturally small, {\em except} in the case where the initial value
of the string coupling is parametrically smaller than the (already very small)
minimal value $ g_{\rm in}^{\min} \simeq 10^{-26}$ needed to solve the horizon
problem, i.e. to generate a space at least as large as our horizon from an 
initial patch of size  $H_{\rm in}^{-1}$\cite{PBB,TW97,LKB00}. 
More precisely, we find
that if we wish generic, coarsely homogeneous, bubbles to evolve into
our (globally very homogeneous) universe we need to require 
 $ g_{\rm in} \,\laq\,  
(10^{-10})^{{\sqrt 3}/2} g_{\rm in}^{\min} \simeq 10^{-35}$.
We note that the necessity (for solving this ``homogeneity problem'') of having
more inflation than the minimal amount needed for solving the horizon problem 
applies also to the standard inflationary models (see below). 

\section{Tensor perturbations in pre-big-bang cosmology}
\label{sec2}
We restrict our investigation to the simplest version of 
the PBB scenario, which is described in the string frame 
by the four-dimensional low-energy string-effective action 
\beq
\Gamma_S = \frac{1}{\lambda_s^2}\, \int d^4 x 
\sqrt{-g_S}\,e^{-\vphi}\,\left [ R(g_S) + g_S^{\mu \nu}\,
\partial_\mu \vphi\,\partial_\nu \vphi \right ]\,,
\eeq
where $\vphi$ is the dilaton field, related to the 
string coupling by $g = e^{\vphi/2}$, and $\lambda_s$ is 
the string scale. In the following, we shall systematically use the string
metric  $g^{\rm S}_{\mu \nu}$ to measure physical lengths or frequencies.
However, it will also be technically useful to introduce the Einstein
metric $g_{\mu \nu}^{\rm E}$.
 The string and Einstein metrics are related (in 4 dimensions)
  by $g^{\rm S}_{\mu \nu} = 
e^{\vphi - \vphi_0}\,g_{\mu \nu}^{\rm E}$. 
Indicating the tensor perturbations as $\delta g_{\mu \nu}^{\rm S} = 
h^{\rm S}_{\mu \nu}$, and working in the synchronous gauge
($g^{\rm S}_{00} = -1,\, g^{\rm S}_{0i} =0,\, g^{\rm S}_{ij} = a^2_{\rm S} \,\delta_{i j}$ 
and $h^{\rm S}_{00} = 0,\, h^{\rm S}_{0 i}=0,\, g_{\rm S}^{i j}\,h^{\rm S}_{i j} =0,\, 
\partial_j h_i^{{\rm S}\,\,j}=0$), 
it is easily checked that $h_{i}^{{\rm S}\,\,j} = h_{i}^{{\rm E}\,\,\,j}$. 
Henceforth, we denote the tensor perturbations 
by $h_{i}^{\,\,j} \equiv h_{i}^{{\rm S}\,\,j} = h_{i}^{{\rm E}\,\,\,j}$. 
Introducing the conformal time $d\eta = dt_{\rm E}/a_{\rm E}=
dt_{\rm S}/a_{\rm S} $ and working in Fourier space we have:
\beq
\label{2.1}
{h_i^{\,j}}(\bx,\eta) = 
\int \frac{d^3 k}{(2 \pi)^3}\,e^{i\bk \cdot\bx}\,
\sum_{\sigma = \pm 2} \epsilon^{(\sigma)\,j}_{i}(\bk)\,h^{(\sigma)}(\bk,\eta)\,,
\eeq
where ${\epsilon^{(\sigma)\,j}}_{i}$ is the polarization tensor, 
which satisfies the usual relations:
\beq
\label{rel}
\epsilon^{(\sigma_1)\,j\,*}_{i}(\bk)\, 
\epsilon^{(\sigma_2)\,j}_{i}(\bk) = \delta^{\sigma_1 \sigma_2}\,, 
\quad \quad 
\sum_{\sigma} \epsilon^{(\sigma)\,j\,*}_{i}(\bk) \, 
\epsilon^{(\sigma)\,l}_{k}(\bk) = \delta^{TT\,jl}_{ ik}(\bk)\,.
\eeq
In the following to ease the notation we shall drop the superscript $\sigma$ 
over $h$ in Eq.~(\ref{2.1}).

\subsection{Evolution of tensor fluctuations} 
\label{subsec2.1}
During the dilaton-driven inflationary (DDI) phase\footnote{Many of the
results below were already derived in \cite{BGGMV95} and other places.
It is, however, simpler to give a self-contained presentation.} 
the Fourier transform of the tensor fluctuations satisfies the equation:
\beq
\label{2.2}
{h}^{\prime \prime} + 2 {\cal H}_{\rm E}\,{h}^\prime + k^2\,{h} = 0\,,
\eeq
where ${\cal H}_{\rm E} = a_{\rm E}^\prime/a_{\rm E}$. Introducing the canonical 
variable $\psi = a_{\rm E}\,h$, we obtain:
\beq
\label{2.3}
{\psi}^{\prime \prime} + 
[k^2 - V(\eta)]\,{\psi} = 0\,,\quad \quad V(\eta) = 
\frac{a_{\rm E}^{\prime \prime}}{a_{\rm E}}\,.
\eeq
{}From the above equation it is straightforward to derive that  
the perturbations propagating inside the horizon 
($k^2 \gg V(\eta)$, i.e. $|k\eta| \gg1$),  
during the DDI phase, evolve simply as 
$ \psi \equiv a_{\rm E}\,h 
\simeq {\rm const.} \times \exp (\pm i k \eta)$, so that (modulo a phase factor)
\beq
\label{2.4}
h_{\rm in\,\, hor.}(k,\eta) = \frac{a_{\rm E}(\eta_{\rm in})}{a_{\rm E}(\eta)}
\,h(k,\eta_{\rm in})\,,
\eeq
where $\eta_{\rm in}$ is some initial time. Note that the scale factor 
$a_{\rm E}$ decreases in time during the DDI era 
[$a_{\rm E} \propto (-\eta)^{1/2}$ and $\eta \rightarrow 0^-$], 
therefore, as long as it is within the horizon,
 $h(k,\eta)$ increases in time during the DDI phase. 
A generic fluctuation exits the horizon for the first time 
during the DDI era at $|\eta_{\rm ex}| \equiv 1/k$. Later on, while 
outside the horizon ($|k\eta| \ll 1$), its evolution is given by
${h}^{\prime \prime} + 2 a_{\rm E}^\prime/a_{\rm E} \,{h}^\prime \simeq 0$,
so that $ h \simeq c_1 + c_2 \int d\eta\, a_{\rm E}^{-2}$.
 As $a_{\rm E}^2 \propto \eta$,
one gets a logarithmic growth
\beq
\label{2.5}
h_{\rm out\,\, hor.}(k,\eta) \simeq \log 
\left ( \frac{\eta}{\eta_{\rm ex}} \right )\,
\frac{a_{\rm E}(\eta_{\rm in})}{a_{\rm E}(\eta_{\rm ex})}\,h(k,\eta_{\rm in})\,.
\eeq
Outside the horizon $h(k,\eta)$ undergoes a logarithmic growth while 
the physical wavelength in the string frame, $\widehat{l}^s = a_S/k$,
is stretched during this DDI phase. In the following we refer 
to physical quantities with a hat, e.g., $\widehat{k}_{i} = k/a_{S\,i}$, 
where $i$ refers to the instant of time $t_i$ at which we evaluate the physical 
quantity.

Later on, if $a_{\rm E}$ starts to increase while the fluctuation is still
outside the horizon the fluctuation $ h \simeq c_1 + c_2 \int d\eta \,
a_{\rm E}^{-2} \simeq {\rm const.}$. 
During the radiation and matter eras the amplitude of 
the tensor fluctuations, after reentering the horizon,
decreases as $\sim 1/a_{\rm E}$, notably as $\sim 1/\eta$ during the 
radiation-dominated (RD) phase and as $\sim 1/\eta^2$ during the matter-dominated (MD) era.

Let us now introduce several (dimensionless) quantities which play a crucial 
role in our analysis: the coefficient ${\cal A}$ which measures the
{\em amplification} of (tensor) fluctuations from the initial time until today,
 the coefficient  ${\cal B} \ll1$ 
(the inverse of the redshift factor) which keeps track of the stretching 
of physical frequencies and lengthscales between the initial patch and now,
and a coefficient ${\cal C}$ whose meaning will be described below :
\beq
\label{2.6}
{\cal A}(\widehat{k}_0) \equiv \frac{h(\widehat{k}_0,\eta_0)}{
h(\widehat{k}_{\rm in}, \eta_{\rm in})}\,, \quad \quad 
{\cal B} \equiv \frac{\widehat{k}_0}{\widehat{k}_{\rm in}}\,,
\quad \quad {\cal C}(\widehat{k}_0) \equiv 
\left (\frac{\hH_{\rm in}}{\hH_{0}} \right )^2\,{\cal B}^2\,
{\cal A}^2(\hk_0)\,.
\eeq
Here, given a comoving wavenumber $k$, $\widehat{k}_0 = k/a_{S 0}$ 
and $\widehat{k}_{\rm in} = k/a_{S \rm in}$.
The index $0$ refers to the present time, $\eta = \eta_0$, while
the index ${\rm in}$ refers to the initial time, $\eta = \eta_{\rm in}$.
The couple of functions of $\widehat{k}_0$, $\{ {\cal A}(\widehat{k}_0),
{\cal C}(\widehat{k}_0) \}$, and the constant ${\cal B}$, exhaust the 
description of the ``transfer function'' between the
classical initial inhomogeneities and the present ones.

For simplicity, we restrict our attention in this paper to the
simplest PBB scenario in which there is not any intermediate phase  
between the DDI era and the standard Friedmann-Lemaitre-Gamow one. 
We denote by $\eta_1$ the conformal time at which 
 the evolution of the Universe (which is always
expanding in the string frame) changes
from the DDI expansion phase to a big-bang fireball. We assume that this
transition takes place when the expansion
rate reaches the string-scale, 
$\hH_1 = \dot{a}_{S 1}/{a}_{S 1} \simeq {\lambda}_s^{-1}$ and when the string 
coupling $g_1=e^{\vphi_1/2}$ 
equals its present value  $g_1 \equiv g_0 \simeq 0.1$.
For times $\eta >  \eta_1$ we assume that the dilaton has
become effectively fixed so that $a_E = a_S$.

If a fluctuation reenters the horizon before the MD era we have:
\beq
\label{2.7}
{\cal A}(\widehat{k}_0) = \frac{a_{\rm E}(\eta_{\rm in})}{a_{\rm E}(\eta_{\rm ex})}\,
\log (k\,\eta_1)\,\frac{a_{\rm E}(\eta_{\rm re})}{a_{\rm E}(\eta_{\rm eq})}\,
\frac{a_{\rm E}(\eta_{\rm eq})}{a_{\rm E}(\eta_0)}\, ,
\eeq
while if reentry occurs after the MD phase, i.e. during the RD phase, we get:
\beq
{\cal A}(\widehat{k}_0) 
= \frac{a_{\rm E}(\eta_{\rm in})}{a_{\rm E}(\eta_{\rm ex})}\,
\log (k\,\eta_1 )\,\frac{a_{\rm E}(\eta_{\rm re})}{a_{\rm E}(\eta_0)}\,,
\label{2.8}
\eeq
where $\eta_{\rm eq}$ stands for the time at which there is equality 
in the Universe between radiation and matter density.

Assuming homogeneity and isotropy, the background fields 
in the string frame evolve as: 
\bea
&& a_S(\eta) = \left (\frac{\eta}{\eta_1}\right )^{-(\sqrt{3}-1)/2}\,, 
\quad \quad \vphi(\eta) = \vphi_1 - \sqrt{3}\log\left (\frac{\eta}{\eta_1}\right )
\quad \quad - \infty < \eta < \eta_1\,, \nonumber \\
&& a_S(\eta) = a_E(\eta)= \left (\frac{\eta}{\eta_1}\right )\,,
\quad \quad \vphi(\eta) = \vphi_1 \quad \quad 
\eta_1 < \eta < \eta_{\rm eq}\,,\nonumber \\
&& a_S(\eta) = a_E(\eta)= \left (\frac{\eta^2}{\eta_1\,\eta_{\rm eq}}\right )\,,
\quad \quad \vphi(\eta) = \vphi_1 
\quad \quad \eta_{\rm eq} < \eta < \eta_{0}\,.
\label{bc}
\eea
Henceforth, to ease the notation, when referring to the 
scale factor in the string frame, we shall drop the 
subscript $S$. Using the above equations, we derive
\bea
\label{2.6b}
&& {\cal A}(\hk_0) 
= \left (\frac{g_1}{g_{\rm in}}\right )\,\left ( \frac{\hH_1}{\hk_{\rm in}} \right )
\,\left ( \frac{\hk_0}{\ho_0^{1}} \right )^{3/2}
\quad \quad \hk_0 \ll  \hH_0\,, \\
\label{2.7b}
&& {\cal A}(\hk_0) 
= \left (\frac{g_1}{g_{\rm in}}\right )\,\left ( \frac{\ho_0^{\rm eq}}{\hk_{\rm in}} \right )
\,\left ( \frac{\ho^1_0}{\ho_0^{\rm eq}} \right )^{1/2}
\,\left ( \frac{\ho^{\rm eq}_0}{\hk_0} \right )^{1/2}\,,
\quad \quad \hH_0 \ll \hk_0 \ll \ho_0^{\rm eq}\,, \\
\label{2.8b}
&& {\cal A}(\hk_0) = \left (\frac{g_1}{g_{\rm in}} \right )\,
\left ( \frac{\ho_0^1}{\hk_{\rm in}} \right )\,
\left ( \frac{\hk_0}{\ho_0^{1}} \right )^{1/2}\,, 
\quad \quad \ho^{\rm eq}_0 \ll \hk_0 \ll \ho^{1}_0\,,
\eea
where $\ho_0^{1} = \omega_1/a_0$, $\ho_0^{\rm eq} = \omega_{\rm eq}/a_0$. 
Here  $\omega_1$ and $\omega_{\rm eq}$  
are the constant comoving wavenumbers whose physical counterparts
 coincide with the 
Hubble expansion rates at time $\eta_{1}$ and $\eta_{\rm eq}$, 
respectively. More explicitly, 
\bea
&& \frac{\omega_{\rm eq}^2}{a_{\rm eq}^2} \equiv \hH_{\rm eq}^2 = 
\frac{8 \pi G}{3}\,\rho_c(t_{\rm eq})\,, \quad \quad 
\frac{\omega_{1}^2}{a_{1}^2} \equiv \hH_{1}^2 = 
\frac{8 \pi G}{3}\,\rho_c(t_{1})\,, \\
&& \frac{\ho^{1}_0}{\hH_1} = \frac{1}{1 + z_{1}} \simeq 10^{-30}\,, 
\quad \quad 
\frac{\ho^{\rm eq}_0}{\hH_0} = \sqrt{1 + z_{\rm eq}} \simeq 10^{2}\,, 
\eea
where we defined the redshift factor $z$ as $a/a_0 \equiv 1/(1 + z)$.
Correspondingly, we get
\beq
{\cal B} = \frac{\hk_0}{\hk_{\rm in}} = 
\left ( \frac{a_{\rm in}}{a_1} \right )\,
\left ( \frac{a_1}{a_0} \right ) = 
\left (\frac{g_{\rm in}}{g_1} \right )^{2/(3+ \sqrt{3})}\, 
\left (\frac{\hH_0}{\hH_1} \right )\,
\left (\frac{\hH_1}{\ho_0^{\rm eq}}\right )^{1/2}.
\eeq
\begin{figure}
\begin{center}
\epsfig{file=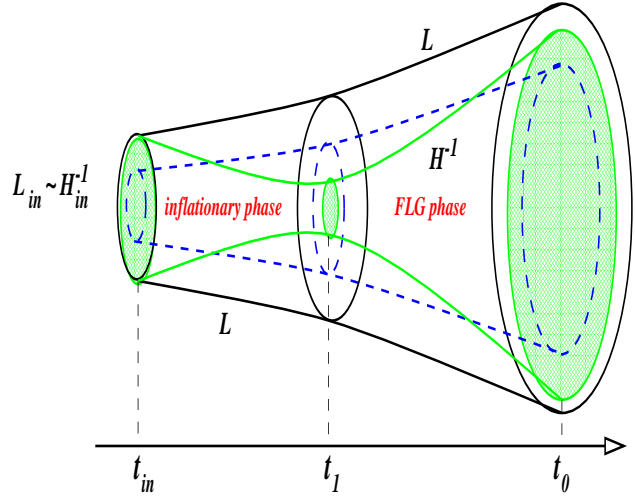,width=0.4\textwidth,height = 0.5\textwidth,angle=-90}
\caption{\sl Schematic representation, 
in the string frame and for the nonminimal version 
of the PBB scenario, of the evolution of: the Hubble 
horizon $H^{-1}$, an intermediate physical wavelength (dashed line) and 
the comoving size $L$ (continuous line) corresponding to the initial patch $H^{-1}$.} 
\label{Fig1}
\end{center}
\end{figure}
It has been derived in Refs.~\cite{TW97,BMUV98} that in order to solve 
the horizon (and flatness) problems in the PBB model, one 
has to require that 
\beq
g_{\rm in} \,\laq \,g^{\rm min}_{\rm in} \simeq 10^{-26}\,, \quad \quad 
\hH_{\rm in}^{-1} \,\gaq \, {(\hH_{\rm in}^{\rm min})}^{-1} \simeq 10^{18}\,
{\lambda}_s\,.
\label{nv}
\eeq
Indeed, defining the total amount of inflation as the ratio 
between the comoving Hubble length at the end and beginning 
of the PBB inflationary phase, 
\beq
{\cal Z} = \frac{a_1\,\hH_1}{a_{\rm in}\,\hH_{\rm in}}\,\,,
\eeq
the horizon problem is solved if we impose that 
\beq 
{\cal Z} \geq \frac{\widehat{l}_0(t_1)}{\widehat{l}_c(t_1)}\,,
\label{hom}
\eeq
where $\widehat{l}_0(t_1) = \hH_0^{-1}\,a_1/a_0$ and 
$\widehat{l}_c(t_1) = \hH_1^{-1} \sim {\lambda}_s$. 
The equality sign in Eq.~(\ref{hom}) refers to the 
{\em minimal} PBB scenario, in which the horizon volume today has evolved 
from an initial (Hubble) patch of size $\hH_{\rm in}^{-1}$. The minimal and 
nonminimal scenarios are illustrated in Fig.~\ref{Fig1}. Note that, 
in the nonminimal scenario the Hubble scale 
at present time $\hH_0^{-1}$ is strictly smaller than the comoving scale
$L(t_0) = L_{\rm in}\,a_0/a_{\rm in}$.
 Using the isotropic and homogeneous PBB background 
solutions (\ref{bc}), it is easily derived that 
$a_{\rm in}/a_1 = (g_{\rm in}/g_1)^{2/(3+\sqrt{3})}$ and 
$\hH_{\rm in}/\hH_1 = (g_{\rm in}/g_1)^{2\sqrt{3}/(3+\sqrt{3})}$.
Imposing Eq.~(\ref{hom}) with the equality sign, 
we find the minimal initial conditions as: 
\beq
g^{\rm min}_{\rm in} \equiv g_1\,
\left ( \frac{\hH_1}{\hH_0} \right )^{-\frac{\sqrt{3}}{2}}\,
\left (\frac{a_1}{a_0} \right )^{-\frac{\sqrt{3}}{2}} = 
g_1\, \left ( \frac{\hH_1}{\ho_0^{\rm eq}} \right )^{-\sqrt{3}/{4}}
\,,
\label{2.9}
\eeq
and 
\beq 
\hH^{\rm min}_{\rm in} 
\equiv \hH_1\,\left ( \frac{\hH_1}{\hH_0} \right )^{-\frac{\sqrt{3}}{\sqrt{3}+1}}
\,\left (\frac{a_1}{a_0} \right )^{-\frac{\sqrt{3}}{\sqrt{3}+1}} = 
\hH_1\,\left (\frac{\ho^{\rm eq}_0}{\hH_1} \right )^{\frac{\sqrt{3}}{2(\sqrt{3}+1)}}\,, 
\label{2.10}
\eeq
where we used $a_1/a_0 = \ho_0^1/\hH_1=(\hH_0/\hH_1)\,(\hH_1/\ho_0^{\rm eq})^{1/2}$.
Inserting in Eqs.~(\ref{2.9}), (\ref{2.10}) the numerical values  
$\hH_0 = 10^{-18}\,{\rm Hz}$, $\ho^{\rm eq}_0 = 10^{-16}\,{\rm Hz}$
and $\hH_1 \simeq {\lambda}_s^{-1} \sim 10^{42}$ Hz, 
we obtain Eq.~(\ref{nv}).

Introducing the notation 
$\og_{\rm in} \equiv g_{\rm in}/g^{\rm min}_{\rm in} \equiv 10^{26}g_{\rm in}$ 
($\og_{\rm in} \leq  1$, with equality in the minimal scenario), we can rewrite 
Eqs.~(\ref{2.7b}), (\ref{2.8b}) in the form:
\bea
\label{2.11}
&& {\cal A}(\hk_0) 
= \frac{1}{(\og_{\rm in})^{1/\sqrt{3}}}\,\left ( \frac{\hH_0}{\hk_0} \right )^{3/2}\,, 
\quad \quad \hH_0 \ll \hk_0 \ll \ho^{\rm eq}_0\,, \\
\label{2.12}
&& {\cal A}(\hk_0) = \frac{1}{(\og_{\rm in})^{1/\sqrt{3}}}\,
\left ( \frac{\hH_0}{\hk_0} \right )^{1/2}\,
\frac{1}{\sqrt{1 + z_{\rm eq}}}\,,
\quad \quad \ho^{\rm eq}_0 \ll \hk_0 \ll \ho^{1}_0\,. 
\eea
The above equations can also be recast in a unique formula which interpolates 
between the two frequency regions:
\beq
\label{2.13}
 {\cal A}(\hk_0) =  \frac{1}{(\og_{\rm in})^{1/\sqrt{3}}}\,
\left ( \frac{\hH_0}{\hk_0} \right )^{1/2}\,\left [
\frac{\hH_0}{\hk_0}  + \frac{1}{\sqrt{1 + z_{\rm eq}}} \right ]\,.
\eeq
The most striking consequence of Eq.~(\ref{2.13}) concerns tensor 
fluctuations on the present horizon scale ${\hk_0} \sim {\hH_0}$.
For these scales, the amplification coefficient connecting them 
to the initial fluctuations is ${\cal A}(\hH_0) \simeq
(\og_{\rm in})^{- 1/\sqrt{3}}$. In the minimal scenario 
($\og_{\rm in} =1$) this is ${\cal A}(\hH_0) \simeq 1$ which means 
that horizon-scale tensor fluctuations today just reproduce (modulo 
a logarithmic amplification factor that we neglected) the corresponding
initial horizon-scale fluctuations.
[Note that this preservation of the amplitude of horizon-scale
fluctuations in the minimal, horizon-solving, case applies equally well to a
standard potential-driven inflation scenario.] On the other hand, the amplification
properties of PBB inflation look worse in the 
nonminimal scenario ($\og_{\rm in} \ll 1$) for which horizon-scale 
tensor fluctuations today are {\em parametrically} amplified compared
to the corresponding initial fluctuations.
This behaviour is different in the PBB model than in
ordinary inflation. For example, if inflation is implemented 
by a de Sitter phase ($\hH_{\rm dS} \simeq {\rm const.}$), 
starting at $t_i$, and ending at $t_1$
when the transition to RD phase occurs, then the amplification factor 
for fluctuations that are just outside the horizon ($\hk_0 \,\sim \,\hH_0$) is:
\beq
\label{ds2}
{\cal A}(\hk_0) = 
e^{-{\cal N}}\,\left (\frac{\hH_{\rm dS}\,a_1}{\hH_o\,a_0} \right )\,,   
\eeq
where ${\cal N} \equiv \log (a_1/a_i)$ is the total number of 
e-foldings. Hence, differently from the PBB scenario, 
where tensor perturbations increase during the PBB era, 
in ordinary inflation as ${\cal N}$ increases (longer inflationary era) 
${\cal A}$ decreases parametrically.

 However, this {\em amplification} of initial
tensor fluctuations by PBB cosmology (instead of the usual 
{\em deamplification} mechanism of potential-driven inflation in the
non-minimal case),
though paradoxical, does not, by itself, imply that the initial
value of the tensor inhomogeneities must be fine-tuned to an
unnaturally small value. Indeed, the classical  quantity that needs
to be smallish for the dilaton-driven inflation to start is not
the {\em amplitude} of tensor waves, but their {\em energy density}
(compared to $\dot \varphi_{\rm in}^2$). We shall postpone the study of
the latter quantity to the next section.

For the quantity ${\cal B}$, defined by Eq.~(\ref{2.6}), we find for  
the PBB scenario under investigation that:
\beq
\label{2.14}
{\cal B} = \frac{\hk_0}{\hk_{\rm in}} = 
\frac{\hH_0}{\hH^{\rm in}_{\rm min}}\,(\og_{\rm in})^{\frac{2}{3 + \sqrt{3}}} 
\simeq 10^{-42}\,(\og_{\rm in})^{\frac{2}{3 + \sqrt{3}}}\,,
\eeq
where in the last equation we used the fact that 
$\hH^{\rm in}_{\rm min} \sim 10^{24}$ Hz.
Here the behaviour is similar to what happens in standard inflation, 
e.g., with a de Sitter phase we derive
\beq
\label{ds1}
{\cal B} = e^{-{\cal N}}\,\left (\frac{a_1}{a_0} \right )\,.
\eeq

In PBB cosmology, as in ordinary inflation,
 the wavelength of the tensor perturbations 
gets always stretched, and the amount of stretching is parametrically larger in the
nonminimal case ($\og_{\rm in} \ll1$; or ${\cal N} > {\cal N}_{\rm min}$ )
 than in the minimal one.

It is important to notice that the formulas given above for the
{\em classical} ``transfer function'' ${\cal A}(\widehat{k}_0)$ and the 
(inverse) redshift factor ${\cal B}$ 
are physically meaningful only when it concerns a present 
spatial frequency $\widehat{k}_0$ such that the
corresponding blueshifted frequency ${\cal B}^{-1}\widehat{k}_0$ (which 
represents the initial frequency) is smaller than the string scale
$\ho_s = 1/{\lambda_s}$. When this is not the case, this classical
transfer function does not apply, and one must consider the problem
of quantum-normalized fluctuations (as studied, e.g., in Ref.~\cite{BGGV95}).
The results for ${\cal A}$ and ${\cal B}$ provided by 
Eqs.~({\ref{2.11}), ({\ref{2.12}) and ({\ref{2.14}) 
are summarized in Fig.~\ref{Fig2}.

\subsection{Power spectrum for tensor fluctuations}
\label{subsec2.2}
The ``bare'' power spectrum is defined by the relation:
\beq
\langle h^{(\sigma_1)}(\bk_1,\eta)\,
h^{(\sigma_2)\,*}(\bk_2,\eta) \rangle = \delta^{\sigma_1 \sigma_2}\,
\delta^{(3)}(\bk_1 - \bk_2)\,{\cal P}_h^{bare}(\bk_1,\eta)\,,
\eeq
where $h^{(\sigma)}(\bk,\eta)$ is given by Eq.~(\ref{2.1}).
\begin{figure}
\begin{center}
\epsfig{file=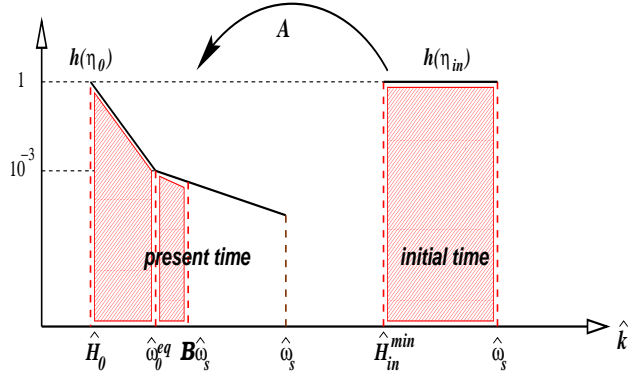,width=0.3\textwidth,height = 0.5\textwidth,angle=-90}
\vspace{0.5cm}
\caption{\sl We show schematically how the ``transfer function'' 
between initial and present time acts both on the amplitude, Eq.~(2.25), 
and the frequency, Eq.~(2.26), of tensor fluctuations. 
We have assumed for simplicity that $h(\eta_{\rm in}) = 1$ 
and $\og_{\rm in} =1$. 
In the minimal PBB scenario the initial string frequency  
$\ho_s = \lambda_s^{-1}$ corresponds today to ${\cal B}\,\ho_s =1\,{\rm Hz}$. 
The present Hubble frequency  
$\hH_0 = {\cal B}\,\hH_{\rm in}^{\rm min} \sim 10^{-18}$ Hz originates from the initial 
frequency $ \hH_{\rm in}^{\rm min} \sim 10^{24}$ Hz, which corresponds to $1$ Fermi.
Note that, whereas the colored regions refer 
to initial classical fluctuations, the white one 
on the left part of the figure concerns wavelengths that would 
correspond to initial length scales formally smaller than the string scale, 
i.e. to quantum fluctuations. For them the classical transfer 
function does not apply.}
\label{Fig2}
\end{center}
\end{figure}
Using the relations (\ref{rel}) and assuming isotropy it is straightforward to derive:
\beq
\langle  h^{\,\,j}_i(\bx_1,\eta)\,h^{\,\,k\,*}_l(\bx_2,\eta) \rangle
= \int \frac{dk}{k}\,\frac{d \Omega_k}{4 \pi}\,
\,e^{i \bk \cdot(\bx_1-\bx_2)}\,
\delta^{TT\,j l}_{i k}(\bk)\,\frac{k^3}{2 \pi^2}\,
{\cal P}_h^{bare}(\bk,\eta) \,. 
\eeq
Hence, the ``physical'' (per logarithmic interval of spatial frequency)
 power spectrum is related to the ``bare'' one through
\beq
{\cal P}_h^{phys.}(\bk,\eta) \equiv \,\frac{k^3}{2 \pi^2}\,
{\cal P}_h^{bare}(\bk,\eta) \,.
\eeq
[In the following we drop the superscript 
``physical'' on the power spectrum.]
Note that the ``physical'' power spectrum has the same dimensions as
 $h^2(\bx,\eta)$ (i.e. it is dimensionless).
The energy density in gravitational waves is given by:
\beq
\rho_{\rm GW} = \int \frac{dk}{k}\,\frac{ d \rho_{\rm GW}(k)}{d \log k}\,, 
\quad \quad  \frac{ d \rho_{\rm GW}(k)}{d \log k} = \frac{1}{16 \pi G}\,
{\cal P}_{\dot{h}}(k)\,,
\eeq
where ${\cal P}_{\dot{h}}$ is the (physical) 
power-spectrum for the time-derivative
of the tensor fluctuation $d h/d t$, i.e. ${\cal P}_{\dot{h}}(k) \simeq \hk^2\,
{\cal P}_{h}(k)$.
The ratio between the energy density in gravitational waves 
and the critical energy $\rho_c$ (conventionally defined in all cases\footnote{
Here, $\hH$,$G$ and $\rho_c$ are all measured in string units. For discussing
initial values it would be more accurate to work with Einstein-frame quantities.
We neglect the inaccuracy (due to the difference between the physical Einstein 
and string Hubble expansion rates) introduced by our definition, which is only 
a factor of order unity.}
by $3\hH^2 = 8 \pi G \rho_c$) reads 
\beq
\Omega_{\rm GW}(\hk) 
\equiv \frac{1}{\rho_{c}}\,\frac{ d \rho_{\rm GW}(\hk)}{d \log \hk}\,
\equiv \frac{1}{6}\,\frac{\hk^2}{\hH^2}\,{\cal P}_{h}(\hk)\,.
\eeq

As explained in the Introduction, in the stochastic PBB model a generic
inflating bubble is expected to have initial values of $\Omega_{\rm GW}(\hk)$
and of similar ratios for the other field inhomogeneities 
which are smallish (say $\sim 1/5$),
but not parametrically small. As we are interested in order-of-magnitude
estimates, we shall henceforth consider that  generic inhomogeneities
should be allowed to be as large as $\Omega_{\rm GW} \sim \Omega_{\vphi} \sim 1$.

Therefore, while by definition the amplification of the power
spectrum of $h$ is given only by the ${\cal A}$ factor 
[defined in Eq.~(\ref{2.6})],
\beq
\frac{{\cal P}_{h}(\hk_0)}{{\cal P}_{h}(\hk_{\rm in})} = {\cal A}^2(\hk_0)\,,
\eeq
the amplification of $\Omega_{\rm GW}$ reads
\beq
\label{Om}
\frac{\Omega_{\rm GW}(\hk_0)}{\Omega_{\rm GW}(\hk_{\rm in})} = 
{\cal C}(\hk_0)\,,
\eeq
where the dimensionless quantity ${\cal C}(\hk_0)$ was defined in Eq.~(\ref{2.6})
above.
We can compute the explicit expression of the tranfer function 
${\cal C}(\hk_0)$ by using Eq.~(\ref{2.13}) and the useful relation 
${\cal B}\,\hH_{\rm in}/\hH_0 = \og_{\rm in}^{2/\sqrt{3}}$. We find:
\beq
{\cal C}(\hk_0) = {\og_{\rm in}}^{2/\sqrt{3}}\,\frac{\hH_0}{\hk_0}\,
\left (\frac{\hH_0}{\hk_0} + \frac{1}{\sqrt{1 + z_{\rm eq}}}\right )^2\,.
\label{n2}
\eeq 
Note the good news that, in this result,
 the power of $\og_{\rm in}$ on the right-hand-side
is now {\em positive} (contrary to the paradoxical negative power of
$\og_{\rm in}$ entering the amplitude-amplification coefficient
${\cal A}(\hk_0)$, given by Eq.~(\ref{2.13})). 
[The positiveness of the exponent of $\og_{\rm in}$ in ${\cal C}(\hk_0)$
is due to the positive compensating exponent entering
${\cal B}\,\hH_{\rm in}/\hH_0 = \og_{\rm in}^{2/\sqrt{3}}$.]
Therefore from the point of view of the parametric dependence of the overall
decrease of tensor inhomogeneities PBB inflation is not qualitatively different
from ordinary inflation. However, we shall see that, from a quantitative point of
view, solving the ``homogeneity problem'' leads to a more severe constraint
for PBB inflation. 

Let us now use the existing limits on the 
amount of gravitational waves generating inhomogeneities  
in the Cosmic Microwave Background Radiation (CMBR) 
to constrain the initial amount of gravitational waves $\Omega_{\rm GW}(\hk_{\rm in})$.
{}From \cite{BA95,M00} we read:
\beq
\label{bound}
\Omega_{\rm GW}(\hk_0)\,h^2_{100} < 7 \times 10^{-11}\,\left ( \frac{\hH_0}{\hk_0} \right )^2\,, 
\quad \quad \hH_0 < \hk_0 < 30\,\hH_0\,.
\eeq
Using $h_{100} = 0.65$ and writing the most stringent consequence of this
limit (corresponding to $\hk_0 = \hH_0$), we get the ``homogeneity constraint'':
\beq
\Omega_{\rm GW}(\hk_{\rm in 0}) (\og_{\rm in})^{2/\sqrt{3}} \alt 10^{-10}\,, 
\label{n1}
\eeq
where we denoted by $\hk_{\rm in 0}$ the initial wavenumber which corresponds
now to $\hH_0$, i.e. $\hk_{\rm in 0} = {\cal B}^{-1} \hH_0$.
[Note that  $\hk_{\rm in 0} \geq \hH_{\rm in}$.]
If we were to restrict ourselves to the minimal PBB scenario
($\og_{\rm in} =1$), i.e. to the case in which 
we require the minimum amount of inflation in order to solve the 
horizon problem 
(i.e. an initial PBB bubble of size 1 Fermi ),
Eq.~(\ref{n1}) would tell us that the initial tensor inhomogeneities must
be unnaturally small: $ \Omega_{\rm GW}(\hH_{\rm in}^{\rm min})  \alt 10^{-10}$.
This would mean that one looses all the genericity benefits of
considering a ``stochastic'' PBB model.

There is, however, a way to solve this ``homogeneity problem'', i.e. to
 relax this unnatural fine-tuning of initial
inhomogeneities, and to allow for  ``generic'' initial inhomogeneities
$ \Omega_{\rm GW}(\hk_{\rm in}) \sim 1$. Indeed, the fact that 
$\og_{\rm in}$ enters Eq.~(\ref{n2}) with a positive power means that
it is enough to impose
\beq
\og_{\rm in} \alt (10^{-10})^{\sqrt{3}/2} \sim 10^{-9}\,.
\eeq
In terms of the string coupling $g_{\rm in}$, this limit is
9 orders of magnitude smaller than the value given in Eq.~(\ref{nv}), 
i.e.
\beq
g_{\rm in} \,\laq\, 10^{-35}\,.
\eeq
Note, however, that this inequality applies only if the initial spectrum is not
completely redshifted out of the present horizon. The condition 
for this is $\hk_{\rm in 0} < {\lambda}^{-1}_s$, i.e.
${\cal B}\,{\lambda}^{-1}_s  > \hH_0$. Using ${\cal B} = 10^{-42}\,
(\og_{\rm in})^{2/(3 + \sqrt{3})}$ this yields:
\beq
g_{\rm in} \,\gaq\, g_{\rm in}^{\rm thr.} = 10^{-69}\,.
\eeq
In conclusion, we obtain three possible scenarios: (i) 
if $10^{-35} \,\laq \,g_{\rm in} \,\laq\, 10^{-26}$, we must require 
initially $\Omega_{\rm GW}^{\rm in} \ll 1 $ and as a consequence, the PBB scenario suffers 
from a serious homogeneity problem; (ii) 
if $10^{-69} \,\laq\, g_{\rm in} \,\laq\, 10^{-35}$,   
there is no need to fine-tune the initial tensor
perturbations,  $\Omega_{\rm GW}^{\rm in} \sim {\cal O}(1)$ 
(in this case, the tensor fluctuations on very large scales can still,
in principle, be seen as classical small fluctuations in the CMBR), and 
(iii)  for $ g_{\rm in} \,\laq\, 10^{-69}$ only initial quantum fluctuations 
survive.

Before ending this section, it is instructive to discuss, for comparison, the fate 
of initial inhomogeneities, discussed so far for the PBB scenario, 
within an ordinary inflation scenario (modelled for simplicity as a simple
de Sitter phase). For a de Sitter 
inflationary phase it is straightforward to derive from 
Eqs.~(\ref{ds2}), (\ref{ds1}) that (for $\hk_0 \,\sim \, \hH_0$): 
\beq
{\cal C}(\hk_0) = e^{-4({\cal N} -{\cal N}_{\rm min})}\,, 
\eeq
where ${\cal N}_{\rm min} = \log ({\hH_{\rm dS}\,a_1}/(\hH_0\,a_0))$ is the 
minimal amount of e-foldings needed to solve the horizon problem.
Applying the CMBR's bound given by Eq.~(\ref{bound}) to these 
fluctuations that are re-entering the horizon now
we get
\beq
\Omega_{\rm GW}^{\rm dS}(\hk_{\rm in 0})\, e^{-4({\cal N} - {\cal N}_{\rm min})} \alt 10^{-10}\,.
\eeq
Therefore, as happens in the minimal PBB scenario,   
in the minimal (horizon-problem-solving)
 de Sitter case (${\cal N} = {\cal N}_{\rm min}$) one is still facing an 
 ``homogeneity problem'', i.e.
the CMBR's bound forces the initial tensor inhomogeneities to be unnaturally small: 
$\Omega_{\rm GW}^{\rm dS}(\hH_{\rm in}^{\rm min})  \alt 10^{-10}$. 
To solve this homogeneity problem, i.e.  to relax this fine tuning and to be able
to start with 
$\Omega_{\rm GW}^{\rm dS}(\hk_{\rm in 0}) \sim 1$, we must depart from 
the minimal de Sitter scenario by at least $6$ e-foldings, i.e. 
${\cal N} \,\gaq \, {\cal N}_{\rm min} + 6$.

\section{Discussion and conclusions}

We have shown that the dilaton-driven inflationary phase of the
 pre-big-bang scenario is not very effective 
in smoothing out the classical inhomogeneities that are
expected to be present in a generic, initial patch of space which
starts its inflationary evolution. We computed the various 
``transfer functions'' which relate the initial spectrum of inhomogeneities
to the present one. Our main conclusion is that the requirement of
naturalness of initial inhomogeneities ($\Omega_{\rm GW} \sim 1$)
can be satisfied only at the price of a constraint \cite{TW97,LKB00} on the
initial value of the (homogeneous part of the) string coupling,
which is much stronger (by a factor $\sim 10^{-9}$) than the previously
acknowledged constraint (following from the necessity to solve
the horizon and flatness problems).

 Ordinary inflation qualitatively faces an analogous homogeneity problem.  
For example in the de Sitter case we need to require $ \sim 6$ e-foldings 
more than the minimal number needed to solve the horizon (and flatness) 
problems in order to overcome this initial inhomogeneity issue.
Quantitatively, this additional constraint is not very severe for ordinary
inflation because, in many inflationary models, the number of e-folds is
exponentially dependent on some inverse power of the coupling constants
of the underlying theory.

This additional ``homogeneity'' constraint on the PBB model discussed here
 does not necessarily
mean that the basic (elegant) idea of dilaton-driven inflation is to be
discarded. There might be other ways of using the kinetic energy
of a scalar field to drive a non-fine-tuned inflationary phase.
In particular the recently proposed model
of ``k-inflation'' \cite{ADM}, which differs from the PBB scenario
 in making use of {\em higher-order} kinetic terms to drive an
 inflationary phase, has been shown to have efficient smoothing 
 properties\cite{GM99}.  

\acknowledgments
We are grateful to Gabriele Veneziano for useful conversations and 
stimulating comments concerning the
homogeneity problem in ordinary inflation.
We thank David Langlois and Maurizio Gasperini 
for useful conversations and/or remarks.
AB's research was supported by the Caltech's Richard Chace Tolman Fund and by
NSF Grant AST-9731698 and NASA Grant NAG5-6840.

\end{document}